\documentclass{INTERSPEECH2023}
\interspeechcameraready 
\usepackage{hyperref}
\usepackage{graphicx}
\usepackage{afterpage}
\usepackage{subcaption}
\usepackage{appendix}
\usepackage{stfloats}
\usepackage{threeparttable} 
\hypersetup{
    colorlinks=true,
    linkcolor=black,
    urlcolor=blue,
    citecolor=black
}
\usepackage{xcolor}
\usepackage{multirow}
\usepackage[normalem]{ulem}

\newcommand{\sous}[1]{\ensuremath{_{\textrm{#1}}}}

\definecolor{suggestion_green}{HTML}{137333}

\renewcommand{\paragraph}[1]{\noindent\textbf{#1}\quad}

\title{Conformer-1: Robust ASR via Large-Scale Semisupervised Bootstrapping}
\name{Kevin Zhang$^*$, Luka Chkhetiani$^*$, Francis McCann Ramirez$^*$, Yash Khare, Andrea Vanzo, Michael Liang, Sergio Ramirez Martin, Gabriel Oexle, Ruben Bousbib, Taufiquzzaman Peyash, Michael Nguyen, Dillon Pulliam, Domenic Donato\thanks{This work was conducted in early 2023. This report was created for internal purposes, and we have made it public on April 12, 2024. The underlying models and algorithms of each ASR provider used in Section 3 may have changed by the time of publication.}}

\address{
  AssemblyAI Inc.}
\email{}

\begin{document}

\maketitle

\begin{abstract}
This paper presents Conformer-1, an end-to-end Automatic Speech Recognition (ASR) model trained on an extensive dataset of 570k hours of speech audio data, 91\% of which was acquired from publicly available sources. To achieve this, we perform Noisy Student Training \cite{park2020improved} after generating pseudo-labels for the unlabeled public data using a strong Conformer RNN-T baseline model. The addition of these pseudo-labeled data results in remarkable improvements in relative Word Error Rate (WER) by 11.5\% and 24.3\% for our asynchronous and realtime models, respectively. Additionally, the model is more robust to background noise owing to the addition of these data. The results obtained in this study demonstrate that the incorporation of pseudo-labeled publicly available data is a highly effective strategy for improving ASR accuracy and noise robustness. 

\end{abstract}

\def\thefootnote{*}\footnotetext{Equal Contribution}\def\thefootnote{\arabic{footnote}}

\section{Introduction}

The power scaling laws introduced in \cite{hoffmann2022training} have shown that the most direct way of improving a Deep Learning model's generalization performance is by scaling both number of parameters in a model and the size of the dataset the model is trained upon. Their work experimentally shows that LLMs (Large Language Models) are usually under-trained and over-scaled, thus proportionally scaling model and dataset size yields rather significant improvements under a smaller compute budget. In the Automatic Speech Recognition (ASR) domain, we usually face a data scarcity issue due to the lack of publicly-available human-labeled speech-to-text datasets. This scarcity has led to much research on how to use the plethora of \emph{unlabeled} audio data that is available.

One research direction uses this unlabeled data for self-supervised pre-training \cite{babu2021xls, baevski2020wav2vec, schneider2019wav2vec}. The central idea in this body of work is to train a powerful speech encoder, such as wav2vec2.0 \cite{baevski2020wav2vec}, in a self-supervised way to extract useful audio features. Once this process is complete, a second training phase commences in which a relatively small quantity of labeled data is used for supervised fine-tuning on the domain-specific task of ASR.

An alternative line of research attempts to address the data scarcity issue head-on. Rather than using large quantities of unlabeled data for \textit{self}-supervised training, this body of work looks to acquire or create large quantities of weakly labeled speech-to-text data. These labels are then used in traditional supervised training schemas. This line of work in turn bifurcates into two main approaches. The first approach relies on generating pseudo-labels using a pre-existing baseline model \cite{park2020improved, kahn2020self, higuchi2021momentum}, while the second approach attempts to source massive amounts of data of ambiguous quality from the public sources and then filter it down to a subset that is both human labeled and high quality \cite{whisper2022}. Our work attempts to address the data scarcity issue head-on and leverages both data filtering and pseudo-labeling to procure high-quality audio and labels at scale.

Following the example provided by \emph{Whisper} \cite{whisper2022}, we sourced audio speech data from open and fair use sources available publicly and  filtered this down to 1.3 million hours. However, rather than filtering out data so that it included high quality human labels, we filtered based on whether the data aligned to the audio domains of our business. This was done to reduce the domain shift our models would be exposed to between training and inference. Since this filtering method does not guarantee labeled audio data, we generated pseudo-labels using a strong Conformer RNN-T baseline model trained on 57k hours of labeled data. Lastly, we adapted the scaling laws proposed in \cite{hoffmann2022training} to the speech domain, and determined the optimal quantity of data to train our model to be 550k hours (Appendix \ref{sec:appendixC}).

Our efforts have resulted in the development of Conformer-1, an automatic speech recognition model trained on 57k hours of high-quality human-labeled and 520k hours of pseudo-labeled speech datasets. Our research has demonstrated that the addition of pseudo-labeled data has led to remarkable relative improvements in Word Error Rate (WER), with our asynchronous and realtime models achieving 11.5\% and 24.3\% relative WER reductions, respectively. As a result, our async model now achieves state-of-art results across various public and internal benchmarks. In order to better approximate human judgment of transcript quality in real-world scenarios, we have introduced a novel Proper Noun metric that complements the more commonly used WER. Through a detailed analysis, we have discovered that the addition of pseudo-labeled public data has significantly improved our model's robustness to noise. Overall, these findings demonstrate the efficacy of Conformer-1 and the potential for pseudo-labeled data to improve ASR performance.
 
\section{Methods}
\subsection{Dataset}
The following section describes the make up of the datasets used to train Conformer-1
\subsubsection{Labeled Dataset}
We use a total of 57k hours of human-transcribed data incorporated from various sources:
\begin{itemize}
    \item \textbf{In-House Data}: Our in-house dataset consists of 51k hours of human-transcribed data. The dataset consists of 20k hours of telephony data acquired from our business, 22k hours obtained from publicly available data sources, and 9k hours of miscellaneous data that consists of privately acquired datasets.
    \item \textbf{Common Voice}: We incorporate 3k hours of data from this Mozilla dataset. \cite{ardila-etal-2020-common}.
    \item \textbf{LibriSpeech}: A corpus of approximately 1k hours of 16kHz read English audiobooks \cite{librispeech}.
    \item \textbf{Synthetic}: We incorporate 2k hours of synthetically generated speech using Multi-Speaker Text to Speech system \textit{VITS} \cite{kim2021conditional}. The goal of this synthetically generated dataset is to expose the model to some meaningful amount of proper noun heavy utterances. For this, we use \textit{News On the Web} \cite{nowcorpus} corpus as a baseline dataset, and extract sentences with normalized distribution of 4 most common proper noun classes \footnote{Most Common Classes: Person, Organization, Location, Date} via \textit{Stanza NER} \cite{qi2020stanza} model. We use \textit{VCTK} \cite{veaux2017cstr} baseline model of \textit{VITS} \cite{kim2021conditional}, and randomly sample one of 110 available speakers for each utterance to maximize speaker variability.
\end{itemize}

\subsubsection{Unlabeled Dataset}

The majority of our training data is composed of audio files acquired from public sources. Despite being extremely valuable, these public datasets lack of human validation, which may lead to low quality audio files (e.g. unnatural words per minute) or simply out of scope (e.g. unsupported languages or categories, music), thereby introducing disruptive noise at training time. In order to overcome such a drawback, filtering techniques can be applied to leave out low-quality audio files and obtain high-quality, in-scope, audio data to be pseudo-labeled.

We carefully select the data sources to match our target domain and language; audio files are further filtered by duration to remove out-of-scope data points.
Secondly, we run all the audio files through Voice Activity Detection (VAD) and our Automatic Language Detection \textit{(ALD)} model, which identifies dominant language of the audio. All the non-English, low-confidence English audios, files with less than 70\% speech activity or with more than 5 seconds of continuous silence, as well as audios where \textit{ALD} and source language mismatch, are removed as a result of this stage.

\paragraph{Pseudo-labeling}
We rely on one of our internal ASR models to generate pseudo-labels for this final filtered unlabeled dataset. We use the model as-is and only change our decoding algorithm to greedy decoding instead of beam search (used in production) as we observe significant computational cost reduction coupled with minimal accuracy degradation during inference.
We then perform a number of postprocessing steps to prepare the data for training. Firstly, audio samples that do not fall within WpM\footnote{WpM - Words per minute} range of $[50, 250]$ are removed. We segment the transcripts and corresponding audio clips into 7 to 20 second clips and filter out segments with a word-level mean confidence value below 0.8: this removes low-confidence pseudo-labels that may introduce noise during training. We also apply additional filtering to avoid replicating known issues that our prior ASR model has exhibited.

\subsection{Model}
We train two types of models currently used in our production environment - \textit{Asynchronous (Async)} and \textit{Streaming (Realtime)}. 

\subsubsection{RNN-T}

Both models use the Transducer model architecture which can be trained with end-to-end RNN-T loss \cite{graves2013speech} using an audio encoder and label predictor. RNN-T loss computes the negative log-likelihood between the predicted and ground-truth transcripts, taking into account all possible alignments between them:

\begin{equation*}\label{eq:1}
    P(y|x) = \sum_{z \in align(y)}P(z|x)
\end{equation*}

Transducer model predicts a probability distribution over the label space at every time step. The probability of an alignment $P(z|x)$ can be factorized as:

\begin{equation*}\label{eq:2}
    P(z|x) = \prod_{i} P(z_{i} | x, t_{i}, \mathrm{Labels}(z_{1:i-1}))
\end{equation*}

where $\mathrm{Labels}(z_{1:i-1})$ is the sequence of non-blank symbols in $z_{1:i-1}$. In our case, $P(z|x)$ is parameterized with a Conformer-based \cite{gulati2020conformer} audio encoder, an LSTM label predictor, and a joint network. The model defines $P(z_{i} | x, t_{i}, \mathrm{Labels}(z_{1:i-1}))$ as:

\begin{gather*}
    \begin{aligned}[t]
        \mathrm{Joint} = \begin{aligned}[t]
            &\mathrm{Linear}(\mathrm{AudioEncoder}(x)_{t_{i}}) + \\
            &\mathrm{Linear}(\mathrm{LabelPred}(\mathrm{Labels}(z_{1:i-1}))) \\
        \end{aligned}
    \end{aligned} \\
    \begin{aligned}[t]
        \qquad P(z_{i} | x, t_{i}, \mathrm{Labels} \begin{aligned}[t]
        &(z_{1:i-1})) = \\
        &\mathrm{Softmax(Linear(tanh(Joint)))} \\ 
        \end{aligned}
    \end{aligned}
\end{gather*}

where $\mathrm{AudioEncoder}(x)_{t_{i}}$ is the encoder output at time $t_{i}$, and $\mathrm{LabelPred}(\mathrm{Labels}(z_{1:i-1}))$ is the label predictor output conditioned on the previous non-blank output sequence. The joint network combines the audio encoder and label decoder outputs using a feed forward layer with softmax output over the vocabulary size.

\subsubsection{Async}
We use Efficient Conformer \cite{burchi2021efficient} as the encoder for the async model. The Efficient Conformer builds on top of the Conformer encoder by introducing progressive downsampling and group attention, resulting in a lowered memory footprint and faster inference compared to its predecessor. We use a larger version of the Efficient Conformer than that reported in the original paper by scaling its hidden dimensions (Table \ref{table:model_hyperparam}).

\begin{table}[t]
\small
\centering
\begin{tabular}{lcc}
\toprule \textbf{Model} & \textbf{Async} & \textbf{Realtime} \\
\midrule
Num Params (M) & 264 & 130\\
Encoder Blocks & 5,5,5 & 17\\
Encoder Dims & 480,768,1080 & 512\\
Attention Heads & 8,8,8 & 8\\
Conv Kernel Size & 15,15,15 & 3\\
Decoder Layers & 2 & 2\\
Decoder Dim & 625 & 512\\ 
\bottomrule
\end{tabular}
\caption{\label{font-table} Model hyper-parameters for Asynchronous Efficient Conformer and Realtime Conformer Transducers.}
\label{table:model_hyperparam}
\end{table}

\subsubsection{Realtime}

\begin{figure}[hbt!]
  \centering
  \includegraphics[width=\linewidth, keepaspectratio]{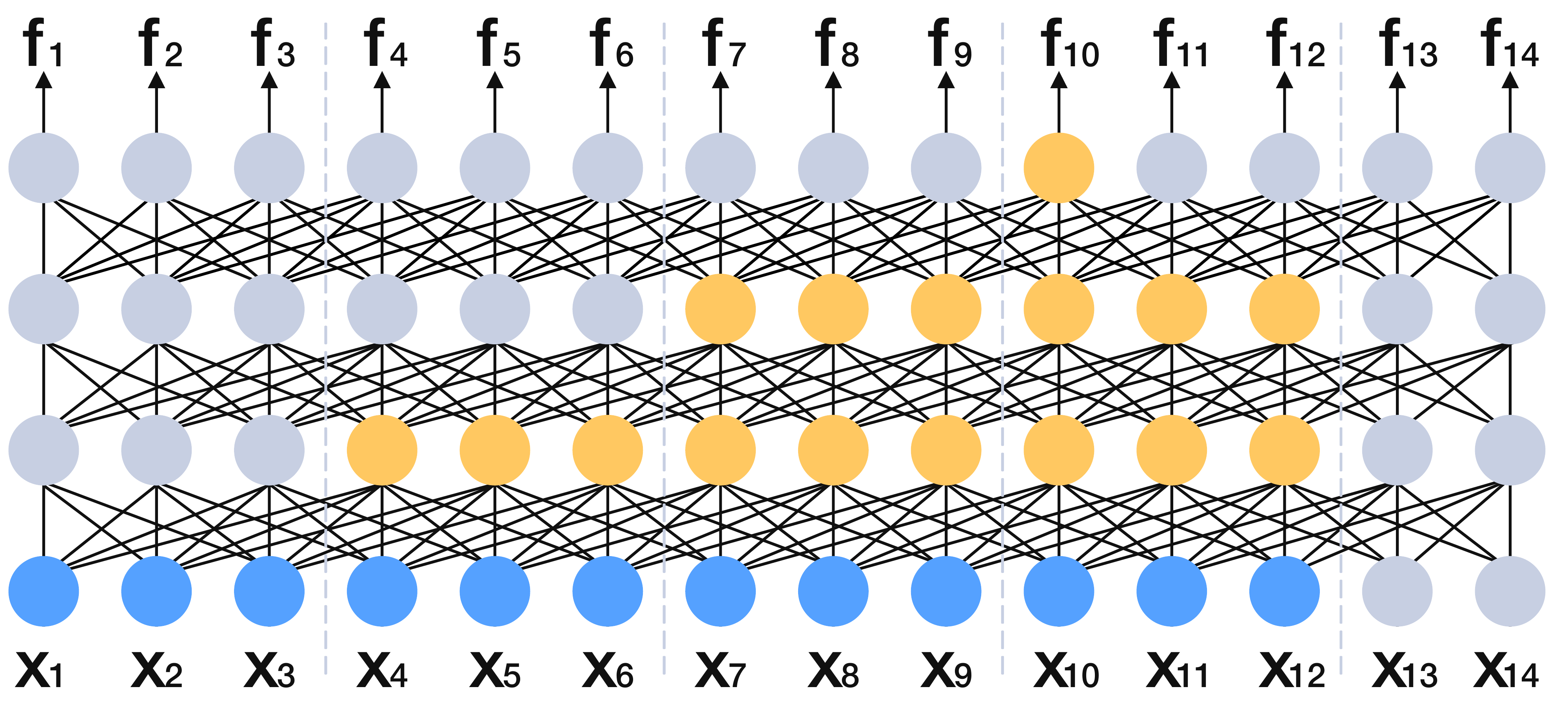}
  \caption{Progressive left-context receptive field of Streaming Transducer, which grows in accordance with depth only on the left side.}
  \label{fig:Realtime Attention}
\end{figure}

For the first step, as a feature extraction and subsampling stage, we use Convolution-based time reduction block. This block consists of 2 Convolutional layers followed by Rectified Linear Unit (ReLU). The subsampling layer is followed by a Conformer \cite{gulati2020conformer} feature encoder, where standard Convolutional layers are substituted by Causal Convolutions \cite{oord2016wavenet} to enable feature caching and streaming.

Following \cite{chen2021developing}, we use chunk-wise diagonal masking strategy during training to hide the future frames from the model's receptive field (Figure \ref{fig:Realtime Attention}). This simple yet effective strategy enables the network to be conditioned on left-context history, thus making it applicable for streaming regime inference. The receptive field is progressively scaled along with the depth of the network.
We limit the number of transformer blocks to 17 and total parameters to 130 million (Table \ref{table:model_hyperparam}) to minimize the response time in the streaming regime.

\subsection{Training}
We use the Adam optimizer \cite{kingma2014adam} with $\beta_1=0.9$, $\beta_2 = 0.999$, $\epsilon=1e-08$, and one cycle scheduler. 
For our async experiments with 520k hours of pseudo-labeled data, we set warmup steps to $21357$, use a global batch size of $1536$ with $16$ per GPU batch size and 4 gradient accumulation steps. For all other experiments, we set warmup steps equivalent to $10\%$ of total iterations, use global batch size of $384$ with $16$ per GPU batch size and $3$ gradient accumulation steps.

For regularization, we apply the same dropout ($0.1$) used in the original Efficient Conformer \cite{burchi2021efficient}; weight decay of $1e-06$; SpecAugment \cite{park2019specaugment} with two frequency masks with mask size parameter $F = 27$, and ten time masks with adaptive size $p_{S} = 0.05$.

We use input spectrograms of 80-dimensional mel-scale log filter banks computed over windows of 20ms, strided by 10ms. We apply WordPiece tokenization \cite{song2020fast} to transform our labels into subword units with vocabulary size of $2048$. The WordPiece tokenizer is trained over the distribution of our training labels. Training is done on NVIDIA A100 80GB GPUs.

\subsection{Inference}
\subsubsection{Async}
Unless otherwise noted, the async model uses beam search with beam size $3$. We integrate an external language model into beam search through shallow fusion \cite{bahdanau2016end} with language model (LM) weight $\lambda$ tuned on the dev-set with grid search. The external LM is an $n$-gram model \cite{brown1993mathematics} with $n = 10$, trained over an external text corpus of 21.5 billion tokens.

The async inference involves several steps. First, we apply the WebRTC Voice Activity Detector (VAD)\footnote{\label{webrtc}https://github.com/wiseman/py-webrtcvad} to identify and remove all silences in the audio. Then, we split the audio into overlapping $25$ seconds chunks. After decoding each and every audio chunk using beam search, we piece them together to construct the transcript for the audio by identifying overlapping sequences in the beginnings and ends of the partial transcripts. We call this technique overlapping inference, and it is motivated by our observation that RNN-T models are more prone to mistakes in the beginning and end of the audio (see Table \ref{table:feature_ablation} for ablation). For inference, we use an exponential moving average (EMA) of model weights maintained throughout training across iterations, with EMA weight set to $9.99999 \times 10^{-1}$.

\subsubsection{Realtime}

As streaming services are required to operate on lower latency constraints compared to asynchronous services, we have made a number of methodically validated choices that result in acceptable performance/quality tradeoff. 
Context window and chunk length constraints are identical during local evaluation and service deployment, thus the results reported in Table \ref{table:pseudo_realtime} are similar for both systems. Specifically, we infer on 450ms chunks by using a greedy decoding strategy to emit the most probable tokens. In contrary to our \textit{asynchronous} service, we don't use Voice Activity Detector (VAD) or Beam Search decoding with external LM shallow fusion. Computations of historical context are cached and amounts to a receptive field with the length of 11 timestamps after subsampling, which approximately equates to 880ms.
Similarly to the asynchronous model, the weights used for inference are EMA (Exponential Moving Average), calculated through our best training run.

\subsection{Evaluation}
\label{sec:eval}
\subsubsection{Dataset}
To evaluate our model's performance, we used a combination of private in-house benchmarks as well as several open source datasets. For our in-house benchmark we acquired data from 6 different domains that we felt provided wide coverage of the majority of industry applications for Speech Recognition: \textbf{broadcasts}, \textbf{podcasts}, \textbf{webinars}, \textbf{telephony}, \textbf{numbers}, and \textbf{generally noisy audio}. These datasets are all human-labeled and acquired through our business operations or obtained from public sources.

For our open source benchmarks, we selected 11 publicly available datasets that are compatible with the APIs of all the providers included in our analysis to avoid introducing sources of error/bias.

\subsubsection{Metrics}
The flagship metric for benchmarking Speech Recognition models is Word Error Rate (WER):
$$ WER = \frac{(Substitutions + Insertions +Deletions)}{Number\ of\ Words\ Spoken}$$
While WER is a valuable metric and has been a staple in research, it is not a holistic measure of performance for ASR.
In particular, a notable issue with WER is that 
it penalizes all types of errors equally. For example, the conjunction ``there's" being transcribed as either ``there is" or ``dare is" incurs the same cost in terms of WER, despite one error being substantially preferable. Issues like these preclude WER from constituting a holistic measure of performance, and models chosen solely on low WER may over-penalize transcriptions that would be deemed correct by human evaluators. In order to remedy the issue we explored various text normalization schemes. The objective of normalization is to remove arbitrary formatting differences between predicted text and labels before calculating WER. These differences include contractions, white space, and filler words (e.g. \textit{ummm}). We settled on the use of OpenAI's text normalizer \cite{whisper2022} before calculating WER to obtain a metric that correlates better with human preferences.

Another important feature in production ASR models is the quality of Proper Noun predictions. Many downstream applications rely on correct and consistent transcription of entities such as names and locations. Despite being a useful quantification of the error in a transcript, WER as a standalone metric fails to capture performance on fine-grained linguistic entities like PNs. As an example, both \textit{Nicholas Cage} and \textit{Ridiculous Cage} would be penalized with a $WER = 0.5$ given the correct spelling \textit{Nicolas Cage}.

To address this shortcoming, we construct a novel metric which we will refer to as \textit{Proper Noun Accuracy}. Operationally, this metric assumes access to two sequences of Named Entities (NEs) which we use as a proxy for Proper Nouns, the former extracted from the target text (or gold), while the latter is obtained from the predicted transcript:

$$gold = \langle (filler_0, \mathtt{type}_0), ..., (filler_n, \mathtt{type}_n) \rangle$$
$$pred = \langle (filler_0, \mathtt{type}_0), ..., (filler_m, \mathtt{type}_m) \rangle$$
where:
\begin{itemize}
    \item $filler$ is the text of the NE (e.g. \textit{Nicolas Cage})
    \item $\mathtt{type}$ is the type of the NE (e.g. \texttt{Person})
    \item $n \neq m$
\end{itemize}

In order to obtain sequences of Proper Nouns in this format we leverage \textit{Named Entity Recognition} (NER) as a proxy to detect and extract Proper Nouns. In particular, we are interested in the Named Entity types \texttt{Person}, \texttt{Organization}, \texttt{GPE} (\textit{geographical entity}), and \texttt{LOC} (\textit{general locations}). We assume access to target and transcript texts that include punctuation and casing and then run these texts through the Stanza NER model \cite{qi2020stanza} to extract the Named Entities. 

Once we have these sequences of Named Entities an alignment stage is necessary to create pairs of ground truth and predicted entities and account for the possible difference in sequence length. Alignment is performed as a two-stage process. We first use \texttt{difflib}\footnote{\url{https://docs.python.org/3/library/difflib.html}} to compute an initial alignment based on the lexical filler. The alignment is then further refined based on the entity type and a threshold of lexical similarity. NEs which do not find their counterpart are not paired with anything and are included as insertions or deletions in the calculation of the final metric. This process provides us with a final list of paired entities in the following form:

$$aligned = \langle (filler_i, filler_j) \rangle, i \in [0, n] \wedge j \in [0, m]$$

Once the alignment is computed, we measure the lexical distance of pairs through a string similarity metric and then average the local scores:
$$PN\_score = \sum_{k=0}^{max(n, m)} = lex\_dist(aligned[k])$$
It is worth noticing that any lexical distance metrics can be used in the final scoring. We chose to apply the Jaro-Winkler edit distance \cite{winkler90} and WER metric to each pair and then average the local scores. Please notice that extracting PNs through Named Entity Recognition also allows characterization of the error by entity type, which can provide a better interpretation of the error itself. When calculating metrics for proper nouns relative to other providers no normalization module was used and the metric was only calculated over datasets where the labels we had were punctuated by a human. We did this because using a normalizer or third party punctation model could bias performance of our chosen NER model. For ablations over different checkpoints of Conformer-1 that include PPN metrics we chose to artificially punctuate and case our internal benchmark datasets labels with Conformer-1's post processing modules, arguing that the bias in this case is consistent. 

In addition, when providing average values over a particular metric for a dataset, we chose to weight metrics for each file by its audio length.

$$ DatasetAverage = \sum_{i=0}^{n} s_i\frac{L_i}{N}$$
Where $N = \sum_{i=0}^{n} L_i$ and $s_i$ is the score for file i. We made this decision after observing that vanilla averaging biases metrics in favor of success on shorter audio files and misrepresents issues that are more prevalent in longer files such as insertions due to noise or hallucinations in periods of silence. 
\section{Experiments}

\subsection{Scaling Pseudo-labels}

\begin{table*}[t]
\centering
\begin{threeparttable}[b]
  \begin{tabular}{l@{\hspace{6pt}}S@{\hspace{5pt}}S@{\hspace{6pt}}S@{\hspace{5pt}}S@{\hspace{6pt}}S@{\hspace{5pt}}S@{\hspace{6pt}}S@{\hspace{5pt}}S@{\hspace{6pt}}S@{\hspace{5pt}}S@{\hspace{6pt}}S@{\hspace{5pt}}S@{\hspace{6pt}}S}
    \toprule
    \multirow{2}{*}{\textbf{Train Data}} &
      \multicolumn{2}{c}{\textbf{Telephony}} &
      \multicolumn{2}{c}{\textbf{Webinar}} &
      \multicolumn{2}{c}{\textbf{Broadcast}} &
      \multicolumn{2}{c}{\textbf{Noisy}} &
      \multicolumn{2}{c}{\textbf{Podcast}} &
      \multicolumn{2}{c}{\textbf{Average}} \\
      & {WER} & {Jaro\sous{ppn}} & {WER} & {Jaro\sous{ppn}} & {WER} & {Jaro\sous{ppn}}
      & {WER} & {Jaro\sous{ppn}} & {WER} & {Jaro\sous{ppn}} & {WER} & {Jaro\sous{ppn}}\\
      \midrule
    57k labeled \textbf{(A)} & 11.0 & 12.4 & 6.5 & 10.5 & 5.2 & 10.3 & 10.7 & 21.5 & 9.9 & 23.2\tnote{1} & 8.7 & 14.4
 \\
    \textbf{A} + 20k pseudo & 11.3 & 12.3 & 6.0 & 11.1 & 4.8 & 10.2 & 10.4 & 21.3 & 9.1 & 14.1 & 8.3 & 13.8 \\
    \textbf{A} + 70k pseudo & 10.8 & 12.1 & 5.7 & 10.5 & 4.6 & 10.2 & 9.9 & 21.4 & 8.5 & 13.3 & 7.9 & 13.5 \\
    \textbf{A} + 120k pseudo & 10.4 & 12.5 & 5.6 & 10.7 & 4.6 & 10.2 & 9.4 & 20.7 & 8.3 & 15.6 & 7.7 & 13.9 \\
    \textbf{A} + 250k pseudo & 10.4 & 12.0 & 5.7 & 10.4 & 4.4 & 10.5 & 9.4 & 21.0 & 8.3 & 13.4 & 7.6 & 13.5 \\    
    \textbf{A} + 520k pseudo & 10.5 & 12.0 & 5.6 & 10.3 & 4.5 & 10.3 & 9.4 & 20.4 & 8.3 & 13.3 & 7.7 & 13.3 \\
    \bottomrule
  \end{tabular}
  \begin{tablenotes}
    \item [1] Jaro-Winkler distance anomaly is caused by misalignments in the named entity detection model.
  \end{tablenotes}  
\caption{\label{font-table} The effects of scaling pseudo-labeled data on async model word error rate (WER) and proper noun accuracy (as measured by Jaro-Winkler distance) across in-house benchmarks. Baseline model is trained on labeled dataset only.}
\label{table:pseudo_async}
\end{threeparttable}
\end{table*}

\begin{table*}[t]
\centering
\begin{threeparttable}[b]
  \begin{tabular}{l@{\hspace{6pt}}S@{\hspace{5pt}}S@{\hspace{6pt}}S@{\hspace{5pt}}S@{\hspace{6pt}}S@{\hspace{5pt}}S@{\hspace{6pt}}S@{\hspace{5pt}}S@{\hspace{6pt}}S@{\hspace{5pt}}S@{\hspace{6pt}}S@{\hspace{5pt}}S@{\hspace{6pt}}S}
    \toprule
    \multirow{2}{*}{\textbf{Train Data}} &
      \multicolumn{2}{c}{\textbf{Telephony}} &
      \multicolumn{2}{c}{\textbf{Webinar}} &
      \multicolumn{2}{c}{\textbf{Broadcast}} &
      \multicolumn{2}{c}{\textbf{Noisy}} &
      \multicolumn{2}{c}{\textbf{Podcast}} &
      \multicolumn{2}{c}{\textbf{Average}} \\
      & {WER} & {Jaro\sous{ppn}} & {WER} & {Jaro\sous{ppn}} & {WER} & {Jaro\sous{ppn}}
      & {WER} & {Jaro\sous{ppn}} & {WER} & {Jaro\sous{ppn}} & {WER} & {Jaro\sous{ppn}}\\
      \midrule
    57k labeled \textbf{(A)} & 13.8 & 13.9 & 9.6 & 12.2 & 7.2 & 11.1 & 15.5 & 26.0 & 13.3 & 15.8 & 11.9 & 15.8 \\
    \textbf{A} + 20k pseudo & 13.3 & 13.5 & 9.0 & 12.4 & 6.1 & 10.9 & 13.9 & 24.4 & 12.3 & 25.4 & 10.9 & 17.3 \\
    \textbf{A} + 270k pseudo & 12.6 & 13.3 & 6.6 & 12.4 & 5.5 & 11.1 & 12.0 & 23.4 & 9.6 & 14.9 & 9.2 & 15.0 \\
    \textbf{A} + 520k pseudo & 12.5 & 12.8 & 6.4 & 12.0 & 5.3 & 11.0 & 11.3 & 23.2 & 9.3 & 14.6 & 9.0 & 14.7 \\
    \bottomrule
  \end{tabular}
\caption{\label{font-table} The effects of scaling pseudo-labeled data on realtime model word error rate (WER) and proper noun accuracy (as measured by Jaro-Winkler distance) across in-house benchmarks. Baseline model is trained on labeled dataset only.}
\label{table:pseudo_realtime}
\end{threeparttable}
\end{table*}

\label{exp:scaling_pseudo}
To investigate the extent to which our ASR model can benefit from pseudo-labeled data, we first train a baseline model on our 57k hours of supervised dataset. Then, we use the baseline model to generate pseudo-labels for up to 520k hours of publicly available data. Finally, we train several models using various amounts of pseudo-labeled data added on top of the original supervised data. These models' performances are reported in Table \ref{table:pseudo_async}.

Overall, introducing additional hours of pseudo-labeled data results in superior performance in both average and proper noun accuracy, although the marginal benefit appears to taper off by 100k hours of pseudo-labeled data. Compared to baseline, introducing 520k additional pseudo-labeled data improves average WER by 11.5\% relative and proper noun Jaro-Winkler distance by 7.6\%. This improvement is consistent across a diverse set of audio domains including telephony, podcast, broadcast, webinar, and noisy audio. Likewise, when we apply the same pseudo-labeled data for training the realtime model, we also observe consistent improvement in accuracy as more pseudo-labeled data is added (table \ref{table:pseudo_realtime}). 

An interesting caveat to note is that, compared to the async model, the realtime model appears to benefit more from the same amount of additional pseudo-labeled data: adding 500k hours of pseudo-labeled data results in a 24.3\% relative accuracy improvement for the realtime model, while only an 11.5\% relative improvement for the async model. In addition, the realtime model appears to exhibit a higher data saturation point - while the benefit of additional pseudo-labeled data tapers off after 100k hours for async, the realtime model still manages to extract value from additional data past 250k hours. On the surface, this seems to contradict the principle that bigger models have a greater data saturation point \cite{hoffmann2022training}. However, we have to keep in mind that both the async and realtime experiments utilize the same pseudo-labels, which are generated by the baseline async model. Since the baseline async model is more accurate than the baseline realtime model, the gap between the quality of the pseudo-labels and the quality of the baseline model is greater in the case of realtime than async, so we can expect the realtime model to extract greater value in the form of accuracy improvement from these pseudo-labels.

Our experiments suggest that the data saturation point for a particular model is sensitive to the quality of the labels. Thus, using more accurate labels can help the training in two ways: (i) to enable the model to converge to a better accuracy given the same amount of data, and (ii) to increase the data saturation point so that the model can benefit from more data. This suggests that a promising way to improve our ASR system further is to find ways to generate better pseudo-labels, using techniques such as ensembling pseudo-labels, better filtering mechanisms, etc.

\subsection{Comparison with other Speech APIs}

\begin{figure*}[hbt!]
    \centering
    \includegraphics[width=\textwidth, keepaspectratio]{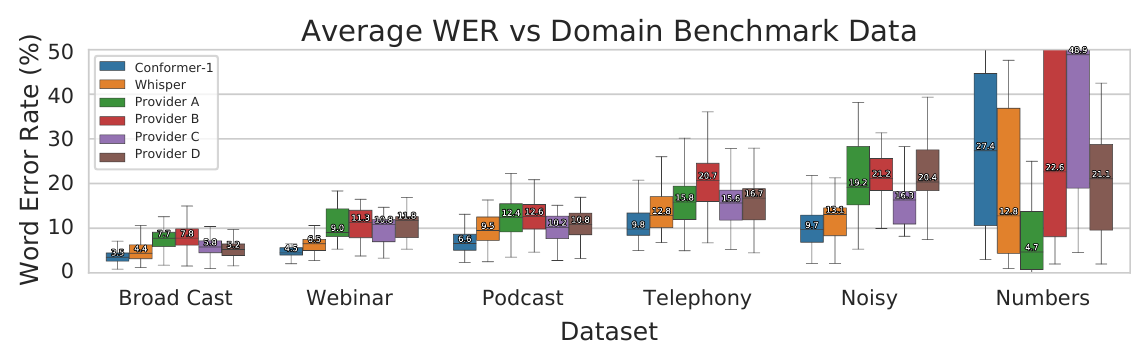} 
    \caption{\textbf{Conformer-1 outperforms Whisper and four other providers in most of our in-house, private benchmarks.}  The distribution of WER from six ASR systems on six in-house, private benchmarks are compared. The boxes show the quartiles of per-example WERs, and the per-dataset aggregate WERs are annotated on each box. The one benchmarks where Conformer-1 does not outperfrom every other model is heavy in numbers, which is an area that we will improve in our next generation of ASR models.}
    \label{fig:limitedheightimage}
\end{figure*}

\begin{figure}[hbt!]
  \centering
  \includegraphics[width=\linewidth, keepaspectratio]{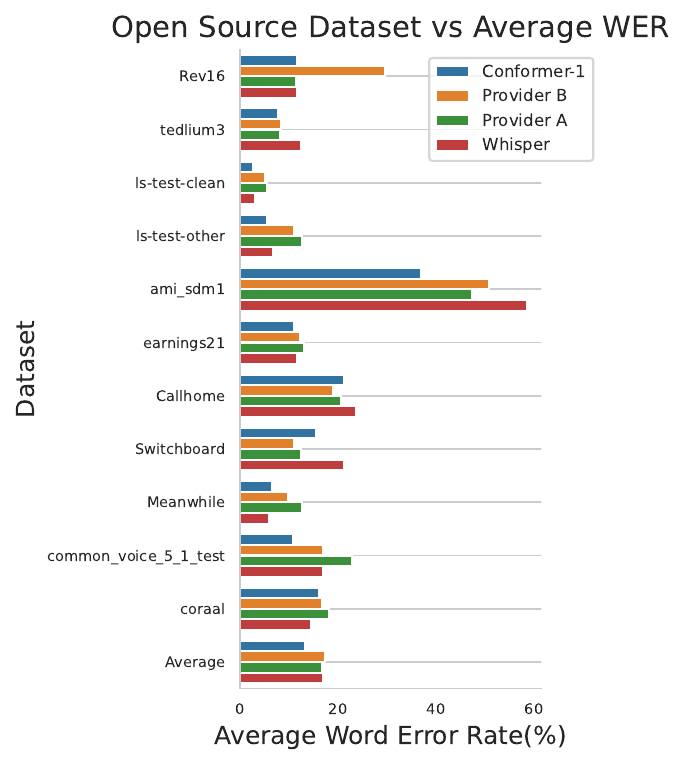}
  \caption{\textbf{Conformer-1 results in lower WER than Whisper for eight out of eleven open source benchmarks.} Conformer-1 also outperforms other two providers in seven out of 11 open source benchmarks.}
  \label{fig:open_source_results}
\end{figure}

\begin{figure}[h]
    \centering
    \subcaptionbox{Plot 1\label{fig:plot1}}{%
        \includegraphics[width=0.45\linewidth]{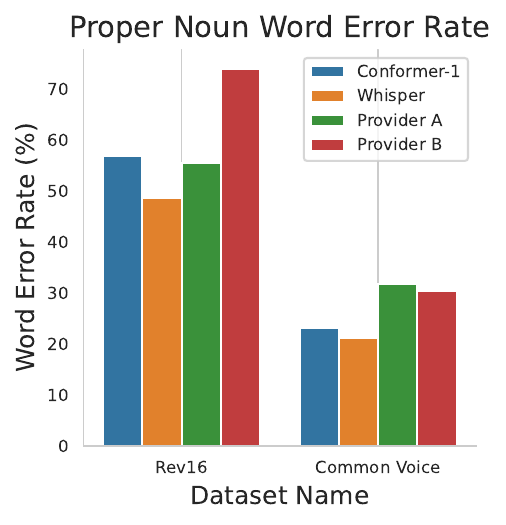}
    }
    \hfill 
    \subcaptionbox{Plot 2\label{fig:plot2}}{%
        \includegraphics[width=0.50\linewidth]{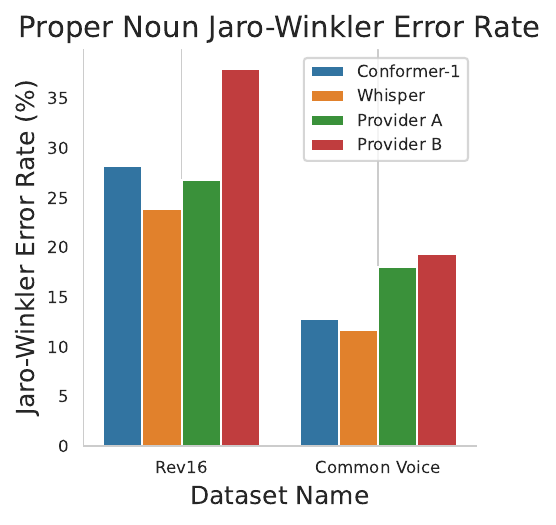}
    }
    \caption{\textbf{Conformer-1 performance on Proper Nouns as measured by WER and Jaro-Winkler metrics.} Only two datasets were used in this analysis because they were the only datasets with punctuated and cased true labels. Conformer-1 does not outperform Whisper on Proper Noun data.}
    \label{fig:side_by_side_plots}
\end{figure}

In order to evaluate our model's performance against other providers we ran the datasets mentioned in Section \ref{sec:eval} through the APIs of providers we identified as being competitive in the space. For our in-house benchmarks we compared against 5 different providers while for the open source benchmarks we narrowed the considered providers to the top 3. To ensure a fair comparison, all commercial ASR services were queried using their default English transcription settings as of March 24th, 2023. The only change we made was to the Whisper API which does not support files over 25 MB. In order to allow transcription of such files, we used a Voice Activity Detection model developed as part of the WebRTC project to find periods of silence over which to split longer files into segments of 5 minutes or less. We then made separate API calls for each chunk and joined the resulting transcriptions together in sequential order. This strategy is in line with what Whisper suggests in their documentation.\footnote{https://platform.openai.com/docs/guides/speech-to-text/supported-languages}

The results of the comparison between our model and commercial providers on in-house benchmarks are shown in Figure~\ref{fig:limitedheightimage}. Conformer-1 outperforms other providers on all domains except \textit{Numbers}: we hypothesize that the difference between Conformer-1 and other models in this domain can be attributed to the use of pseudo-labels and our filtering strategy. We leave this discrepancy open as a future area of research.

The results of the comparison performed on open source datasets are summarized by Figure \ref{fig:open_source_results}. While the results are less pronounced than on our in-house benchmarks, Conformer-1 is still competitive or wins outright on every benchmark tested. We note that while we were able to control for unseen data with our proprietary benchmarks datasets, since the open source benchmarks are public some of these test sets could have been included in the providers training datasets and misrepresent the robustness their models.

In addition, we were able to run a limited analysis of Conformer-1's performance on proper nouns on datasets that provide human punctuated and cased labels. Although constrained, these results summarized in Figure \ref{fig:side_by_side_plots} show that despite the use of noisier data in the form of pseudo-labels, Conformer-1 remains within a competitive range of other providers on Proper Noun Accuracy. We consider this result preliminary and plan to conduct a more thorough analysis in follow-up work.

\subsection{Robustness to Noise}

\begin{figure}[h]
    \centering
    \subcaptionbox{Plot 1\label{fig:plot1}}{%
        \includegraphics[width=0.48\linewidth]{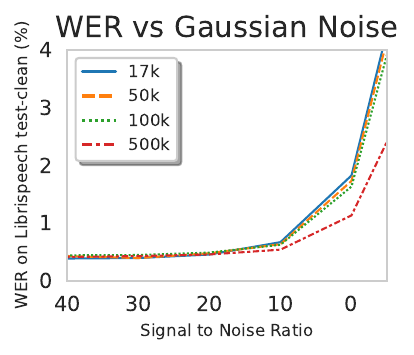}
    }
    \hfill 
    \subcaptionbox{Plot 2\label{fig:plot2}}{%
        \includegraphics[width=0.48\linewidth]{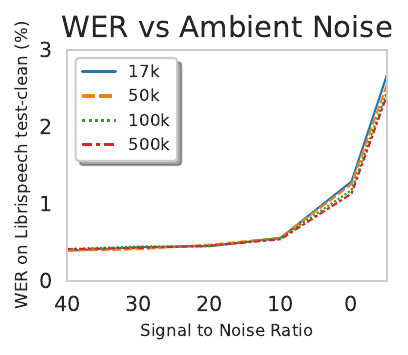}
    }
    \caption{Noise robustness with increase in amount of pseudo labeled data}
    \label{fig:side_by_side_plots_1}
\end{figure}

\begin{figure}[h]
    \centering
    \subcaptionbox{Plot 1\label{fig:plot1}}{%
        \includegraphics[width=0.45\linewidth]{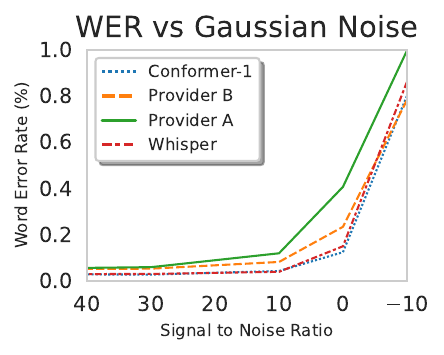}
    }
    \hfill 
    \subcaptionbox{Plot 2\label{fig:plot2}}{%
        \includegraphics[width=0.45\linewidth]{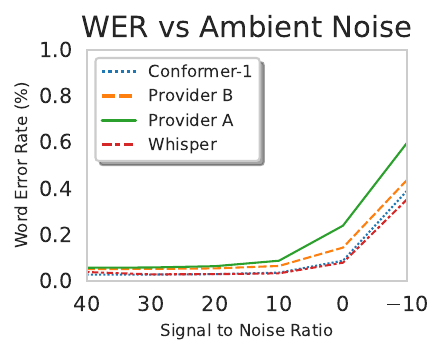}
    }
    \caption{The effect of Gaussian noise and Ambient noise on WER on Conformer-1, Whisper, and two other providers.}
    \label{fig:gaussian_noise_wer}
\end{figure}
In order to test our model's ability to handle noisier data we conducted ablations applying varying levels of noise to the \textit{LS-test-clean} dataset \cite{librispeech}. We took the audios from these datasets and injected clean LibriSpeech audio with either white Gaussian noise or background noises from the Microsoft DNS dataset \cite{dubey2023icassp}. For the Gaussian noise we sample noise from a Gaussian distribution at various SNRs. For the background noise experiment, a noise file from the DNS dataset is randomly selected and combined with the clean LibriSpeech file at varying SNR's.

We hypothesize that as the size of our training dataset increased we would see corresponding gains in performance on these noised files. To test this, we performed ablations on model variants trained on a varying amounts of data. We used four models trained on 57k hours of human labeled speech along with increasing amounts of pseudo-labeled data, 17k hours, 50k hours, 100k hours and 500k hours. In order to isolate the source of increased noise, all experiments for this ablation used deterministic greedy decoding strategies as opposed to the beam-search strategies employed in commercial APIs. Figure \ref{fig:side_by_side_plots} shows that as the amount of training data increases, the models show smaller increases in WER with increasing noise. It should be noted that this increased robustness is much more present for the audio augmented with Gaussian noise. While still there the gains in robustness are smaller for ambient noise, suggesting that our model could benefit from the addition of these types of noises during training. We think Gaussian noise more closely resembles the differences in audio quality present in the public data sources and attribute this result to that.  These results show that scaling noisier pseudolabeled data leads to corresponding increases in  a models robustness to noise.

Figure \ref{fig:gaussian_noise_wer} shows how the performance of various providers decays as the SNR is increased for both Gaussian noise and background noise. There is a sizeable gap between Conformer1/Whisper and the other tested providers. For the Gaussian noise ablation, Conformer-1 has a slight edge over Whisper while performing slightly worse for ambient noise. Based on our above analysis, we attribute this result to the sheer size of the datasets that both models were trained on, covering a much wider distribution than other available models. These results demonstrate our model's versatility and robustness under changing data distributions.

\section{Related Work}
In recent years, several methods have been explored to enhance the performance and generalizability of Automatic Speech Recognition (ASR) models by leveraging unsupervised and/or weakly supervised datasets.

In \cite{park2020improved}, the authors propose an iterative training approach that utilises a predominantly imbalanced set of labeled and unlabeled datasets. The proposed technique entails fine-tuning a base model on a small labeled dataset, followed by generating pseudo-labels on a larger dataset using external language model rescoring and gradational filtering. The newly initialized model is trained on the combined human-labeled and pseudo-labeled datasets and subsequently used to regenerate pseudo-labels as it converges to a better \textit{Word Error Rate}. This process of iterative pseudo-labeling, filtering, and re-training is repeated until the model signals convergence. The authors use a 960-hour human-annotated version of the \textit{LibriSpeech} dataset \cite{panayotov2015librispeech}, as well as 60,000-hour unsupervised set of \textit{Libri Light} \cite{kahn2020libri} for iterative pseudo-labeling. With the Noisy Student training approach, authors were able to improve upon the previous semi and fully supervised state of the art results on LibriSpeech \cite{panayotov2015librispeech} 100 hour and 960 splits.

In a follow-up research work \cite{kahn2020self}, authors employ \textit{wav2vec 2.0} \cite{baevski2020wav2vec} pre-training method to further improve the robustness of the base model. Other works employ alternative base datasets \cite{panayotov2015librispeech, kahn2020libri} and training methods \cite{park2020improved} for iterative pseudo-labeling.

Authors in \cite{higuchi2021momentum} take pseudo-labeling further by generating pseudo-labels on the fly, removing the need for iterative manual process. Inspired by the \textit{Mean Teacher method} \cite{tarvainen2017mean}, authors conduct the training process by first training a seed model on the supervised dataset. Afterwards, an online student and offline teacher are both initialized using the copies of the seed model. During training, the student learns from the pseudo-labels generated by the teacher, while the teacher maintains an exponential moving average of the student weights. The training is continued without external intervention by this online-offline fashion until convergence. The authors argue that this collaborative method of \textit{``dual training''} introduces robustness to the student network as the implementation of momentum update facilitates a more seamless evolution of the offline model.

\textit{Whisper} \cite{whisper2022} presents an encoder-decoder based Transformer \cite{vaswani2017attention} model for multi-task, multi-language speech modeling tasks, which achieved state-of-the-art results surpassing its predecessors on various downstream tasks including Speech Recognition, Language Identification, and Speech Translation. The authors leverage a large-scale weakly supervised dataset, consisting of a total of 680k hours. The dataset is rigorously filtered and formatted to improve the quality of the weakly supervised data. By scaling the model size and the percentage of dataset used, the authors show experimentally that the model is able to converge to lower error rates across various tasks and datasets. One of the key findings in \cite{whisper2022} is that large-scale weakly supervised training enables the model to be more consistent across various domains and datasets. Specifically, when benchmarked on the thirteen most popular English Speech Recognition datasets, the largest proposed model \textit{Whisper Large V2}\footnote{Whisper Large V2 is 1.55 billion parameter model trained on full-scale dataset} results in a standard deviation of 9.3 in Word Error Rate, compared to the twice-as-large 18.9 of wav2vec 2.0 \cite{baevski2020wav2vec}.

Our goal is to build on top of these recent innovations to pick the most applicable and resource-friendly recipes. More specifically, we train a strong baseline model on a large human-labeled dataset and utilize pseudo-labeling to scale and diversify our dataset up to 570k hours in total. In contrast with \cite{park2020improved}, we apply greedy decoding to generate our pseudo-labels and only do one iteration of noisy student training to make our training feasible at scale. Compared to \cite{whisper2022}, we use a confidence-based filtering strategy using the average word confidence acquired from greedy decoding emissions to increase the quality of our pseudo-labels. Our choice of \textit{number of parameters} versus \textit{hours of dataset} builds on the work of \cite{hoffmann2022training} to optimize the tradeoff between inference time, training time, and accuracy.\footnote{By applying the scaling formula we arrived at 550k hours of dataset for 264 million parameter model.}

This work encouraged us to explore the relationship between parameter count and data size for the task of Speech Recognition, motivating a focus on determining benefits of the data-only scaling strategy. We experimentally show that scaling the dataset in accordance with the scaling laws proposed by \cite{hoffmann2022training} (Appendix \ref{sec:appendixC}) while keeping the model size fixed provides sizeable improvements in terms of accuracy and robustness.

\section{Limitations and Future Work}

In this paper we replicated the success other papers have had by obtaining weakly supervised data from the public sources with much less resources by simply pseudo-labeling unsupervised data. However this conclusion is purely based off empirical results and further theoretical analysis of these results is an open area of exploration. We hypothesize that pseudo-labels are helping for different reasons, including suppressing the negative effects of outlier samples and covering a wider train distribution, but want to get a more scientific basis for our explanations. In future work we would like to control different elements of our datasets to better understand how pseudo-labels are changing our training distribution.

Given that the quality of pseudo-labels plays an important role in training, we would also like to experiment with more sophisticated pseudo-labeling strategies including ensembling the pseudo-labels over different models, applying temperature sampling during pseudo-labeling to achieve greater diversity, and using better filtering methods. 

Moreover, our analysis of performance on named entities is limited by the difficulty of obtaining punctuated and cased labels. In order to avoid introducing bias into our analysis by having to artificially punctuate and case labels, we restricted our proper noun analysis only to datasets that had punctuation and casing out of the box, which ended up being quite an aggressive filter. In follow-up analysis, we would like to get our in-house benchmarks cased and punctuated by human annotators to get a more thorough comparison of our model's performance on named entities metrics relative to other providers. 

We also show evidence that our overlapping decoding strategy is beneficial to the performance of transducer models. Future research areas to support this would involve confirming these results on other open source baselines.

Lastly, Conformer-1's output had to be post-processed to add punctuation, casing, and ITN. This potentially introduces additional sources of error, and we hypothesize that this could be part of the reason that Whisper and other providers perform slightly better in proper noun accuracy. Our future work will attempt to address this by training end-to-end ASR models with punctuation and casing integrated into the output.

\section{Conclusion}
Conformer-1 achieves state-of-the-art performance in terms of WER on a diverse set of open-source and private in-house benchmarks; in particular, it outperforms Whisper in eight out of eleven open-source benchmarks. Given these results, we show that it is possible to train state of the art Speech Recognition systems without relying on large amounts of human-labeled data through resource intensive methods. In addition, we provide evidence that scaling up pseudo-labeled data for diverse audio such as publicly available sources can lead to a corresponding increase in a model's ability to maintain performance in noisy environments. Lastly, we have proposed a novel metric to quantify Proper Noun Accuracy, in the hope that it will lead to an increased emphasis within the field on the inclusion of punctuation and casing in benchmark datasets and the development of metrics more correlated with human judgement. 

\section{Acknowledgments and Statement of Contributions}
Takuya Yoshioka, Ryan O'Connor and Marco Ramponi for reviewing and helpful discussions.

\newpage
\bibliographystyle{IEEEtran}
\bibliography{mybib}

\clearpage %
\appendix
\begin{appendices} 

\section{The Effect of Stochasticity on Bootstrapped Learning}

Intuitively, learning from a model's own outputs (pseudo-labels) should not improve its performance, since the $p(output|input)$ for the pseudo-labels and the model is identical by definition, so in a way the model is not being offered anything new to learn from. In an attempt to explain why pseudo-labeling helps, we hypothesize that the stochastic elements in the training such as SpecAugment \cite{park2019specaugment} and dropout \cite{srivastava2014dropout} are necessary for the model to learn from its own pseudo-labels, and that they do so by perturbing the input distribution, which creates a gap between $p(output_{pseudo}|input)$ and $p(output_{model}|input)$, encouraging the model to learn beyond it's original distribution. To test this hypothesis, we run two experiments where we finetune a seed baseline model for 1 epoch on 7.3k hours of audio pseudo-labeled with the same baseline model with and without stochastic features in the training pipeline (``stochastic" and ``deterministic" model variant respectively). Specifically, we remove dropout and SpecAugment for the deterministic variant and keep the stochastic variant's hyperparameters unchanged from Conformer-1 training settings. If our hypothesis was valid, then the deterministic experiment should not benefit from further finetuning on the pseudo-labeled data.

\begin{table}[t]
\small
\centering
\begin{tabular}{lccc}
\toprule \textbf{Dataset} & \textbf{Baseline} & \textbf{Deterministic} & \textbf{Stochastic}\\
\midrule
Telephony & 11.5 & 12.0 & 11.1 \\
Webinar & 6.3 & 6.1 & 6.0 \\
Broadcast & 4.9 & 5.0 & 4.9 \\
Noisy & 10.5 & 10.9 & 10.5 \\
Audio & 9.3 & 9.4 & 9.2 \\
Numbers & 6.4 & 5.7 & 4.9 \\
Average & 8.1 & 8.2 & 7.8 \\
\bottomrule
\end{tabular}
\caption{\label{font-table} In-house WER benchmarks on our test set for "stochastic",  "determinstic" and baseline V9 checkpoint variants.}
\label{table:stochasticity}
\end{table}

Our results shown in \ref{table:stochasticity} supports this hypothesis. Finetuning on pseudo-labels in the absence of stochasticity results in slight degradation in WER, possibly due to over fitting to the baseline distribution. On the other hand, the stochastic variant performs better than both the baseline and the deterministic variant, validating our original hypothesis that the addition of stochasticity is necessary for the model to learn from its own pseudo-labels.

\section{Decoding Strategies for Long-form Transcription}
In table \ref{table:feature_ablation}, we show the effect of decoding features on model accuracy. First, we show that using voice activity detector (VAD) to strip out silences prior to decoding has no effect on accuracy, which indicates that the model already does well in the presence of silences (i.e. does not produce hallucinations). Regardless, we keep VAD in the inference pipeline because it can potentially reduce the processing time of audio with long silences by cutting them out. Second, we show that overlapping inference significantly improves the model accuracy. From our observations, RNN-T models perform poorly near the beginning and end of the utterances. Overlapping inference resolves this issue by decoding overlapping audio segments and joining the overlapping partial transcripts by matching strings near the ends. Thirdly, applying beam search does not seem to affect average accuracy, but can benefit proper noun accuracy slightly. Due to the rarity of proper nouns in the training data, the model is likely to experience greater ambiguity for predicting the correct tokens for a proper noun, meaning that the conditional probabilities for proper noun tokens have higher entropy. Thus, an algorithm like beam search that can explore a larger space of possible sequences is expected to do better in this context. Lastly, integrating external LM does not appear to improve average or proper noun accuracy. The likely explanation is that the benefit of external LM decreases as the model is trained on increasing amount of audio data, as the distribution gap between the labels in audio training data and LM training data narrows.

\begin{table}[t]
\small
\centering
\begin{tabular}{lcc}
\toprule \textbf{Features} & \textbf{WER\sous{avg}} & \textbf{Jaro\sous{avg}}\\
\midrule
Greedy & 8.2 & 13.5  \\
+ VAD & 8.2 & 13.6   \\
+ Overlapping Inference & 7.6 & 13.5  \\
+ Beam Search & 7.6 & 13.3  \\
+ External LM & 7.7 & 13.3 \\
\bottomrule
\end{tabular}
\caption{\label{font-table} The effects of adding decoding features on average WER and average proper noun accuracy (as measured by Jaro-Winkler distance) across in-house benchmarks.}
\label{table:feature_ablation}
\end{table}

\section{Derivation of Optimal Train Hours}
\label{sec:appendixC}
According to the \textit{Table A3} from \cite{hoffmann2022training}, the optimal token count to number of non-embedding parameters ratio is roughly 20x. To adapt this simple formula to \textit{Speech Recognition} use-case, we have made a number of assumptions, namely:
\begin{itemize}
\item{Average word count of standard English speaker is 120 words per minute (\textit{WPM})}.
\item{Tokens\footnote{WordPiece tokenization \cite{song2020fast} with vocabulary size of 2048} per word ratio is $\frac{4}{3}$ (\textit{TPW})}.
\item{Required tokens per model parameter (\textit{TPP}) is 20x.}
\end{itemize}
Thus, the formula for calculating number of required hours of speech data is simply as follows:
$$ 
total \ hours = \frac{Parameters \cdot TPP}{WPM \cdot TPW \cdot 60}
$$

The estimated dataset size for our 264M parameter asynchronous model (Table {\ref{table:model_hyperparam}}) is 550k hours.
\end{appendices}
\end{document}